\documentclass[twocolumn,aps,prl,amsmath,amssymb,superscriptaddress,floatfix]{revtex4-1}

\usepackage[caption=false]{subfig}
\usepackage[percent]{overpic}
\usepackage{graphicx}
\usepackage{dcolumn}
\usepackage{bm}
\usepackage{multirow}
\usepackage{amssymb}
\usepackage{amsmath}
\usepackage{xcolor}
\usepackage{braket}
\usepackage{CJK}
\usepackage{indentfirst}
\usepackage{cases}
\usepackage{placeins}
\usepackage[colorlinks,citecolor=blue,linkcolor=red,urlcolor=blue]{hyperref}
\newcommand{\PRLsection}[1]{{\it #1} --- }

\begin{document}

\title{Metallic Gross-Neveu criticality and superconductivity on the \texorpdfstring{$\mathrm{SO}(3)$}{SO(3)} SLAC fermion}

\author{Feng-Yu Zhang}
    \altaffiliation{These authors contributed equally to this work.}
    \affiliation{Guangdong Provincial Key Laboratory of Magnetoelectric Physics and Devices, School of Physics, Sun Yat-sen University, Guangzhou 510275, China}
    \affiliation{School of Physics, Sun Yat-sen University, Guangzhou 510275, China}
\author{Yin-Kai Yu}
    \altaffiliation{These authors contributed equally to this work.}
    \affiliation{Beijing National Laboratory for Condensed Matter Physics \& Institute of Physics, Chinese Academy of Sciences, Beijing 100190, China}
    \affiliation{University of Chinese Academy of Sciences, Beijing 100049, China}
\author{Zi-Xiang Li}
    \email{zixiangli@iphy.ac.cn}
    \affiliation{Beijing National Laboratory for Condensed Matter Physics \& Institute of Physics, Chinese Academy of Sciences, Beijing 100190, China}
    \affiliation{University of Chinese Academy of Sciences, Beijing 100049, China}
\author{Shuai Yin}
    \email{yinsh6@mail.sysu.edu.cn}
    \affiliation{Guangdong Provincial Key Laboratory of Magnetoelectric Physics and Devices, School of Physics, Sun Yat-sen University, Guangzhou 510275, China}
    \affiliation{School of Physics, Sun Yat-sen University, Guangzhou 510275, China}

\date{\today}

\begin{abstract}
    The realization of Dirac criticality beyond the conventional Gross-Neveu-Yukawa (GNY) paradigm has become a major frontier in condensed matter physics. In this work, we introduce an $\mathrm{SO}(3)$-symmetric bilayer SLAC fermion model with tunable inter-layer interactions that exhibits a rich quantum phase diagram. As the interaction strength increases, the system undergoes two distinct phase transitions. The primary transition is a continuous boundary separating a Dirac semimetal (DSM) from an $\mathrm{SO}(3)$-broken ordered phase. Crucially, this transition evades the standard GNY universality class because the emergent order only gaps out a subset of the itinerant fermions. Using large-scale quantum Monte Carlo (QMC) simulations, we establish that this transition belongs to the Gross-Neveu-$\mathrm{SO}(3)$ universality class with $N=6$ irreducible Dirac cones and precisely extract the corresponding critical exponents. At stronger couplings, a second transition drives the system into an inter-layer $\mathrm{SO}(3)$-symmetric superconducting (SC) state. We provide strong numerical evidence that this transition is first-order. Our study provides new insights into the exploration of Dirac criticality beyond the standard GNY universality class, and also offers a distinct platform for investigating $\mathrm{SO}(3)$-symmetric superconductivity.
\end{abstract}

\maketitle

\PRLsection{Introduction}
    Quantum phase transitions in Dirac fermion systems represent a central research frontier in modern physics~\cite{Sachdevbook}, with early investigations originating in high-energy physics~\cite{Gross1974prd}. In recent years, inspired by groundbreaking progress in graphene~\cite{Geim2009rmp} and topological quantum materials~\cite{KaneReview,SCZhangReview}, research on quantum criticality of Dirac fermions has attracted extensive and growing attention in condensed matter physics~\cite{Sorella1992,Herbut2006prl,Herbut2009prb,Herbut2013prx,Sorella2016prx,Sherer2017prb,Sherer2017prd,Sheng2014science,Seifert2020prl,Janssen2021prb,Corboz2018prx,Andreas2018prx,Qi2022PRB,Wang2014NJP,Li2015NJP,Guo2021PRB,Assaad2022PRL,Janssen2014prb,Knorr2016prb,Gracey2016prd,Roy2016JHEP,Gies2015prd,Sherer2017prb,Sherer2017prd,Janssen2023prb,Scherer2018prb,Joseph2021review,Roy2016PRB,Moon2018PRB,Poland2019rmp,Corboz2018prx,Andreas2018prx,Lang2019PRL,Liuzh2023prl,Li2018SciAdv,Vishvanath2017NP,Wu2016PRB,Vaezi2022PRL,Sorella2018PRB,Xu2021PRL,Wang2026Honeycomb,Meng2020PRB,Chandrasekharan2013PRD,Guo2022PRB,Meng2023arXiv,Wang2023PRR,Yu2024PRL,Li2024prl,Hohenadler2019PRL,Roy2016PRB,Moon2018PRB,Scalettar2019PRL,Herbut2023arXiv}.

    It has been well established that gapless fluctuations of Dirac fermions greatly enrich the fundamental understanding of quantum critical phenomena and give rise to the Gross-Neveu universality class. In the conventional Gross-Neveu universality class, all Dirac cones become simultaneously gapped when the bosonic field spontaneously breaks the associated symmetry. This picture not only provides a mechanism for fermion mass generation in high-energy physics~\cite{Gross1974prd}, but also accounts for the origin of numerous ordered phases in condensed matter systems~\cite{Sorella1992,Herbut2006prl,Herbut2009prb,Sorella2016prx,Assaad2022PRL}.

    Phase transitions in Dirac fermion systems beyond the standard Gross-Neveu (GN) universality class are therefore of considerable interest. Recently, a novel unconventional GN-$\mathrm{SO}(3)$ universality class has been proposed~\cite{Seifert2020prl}. This universality class governs the transition between a symmetric Dirac semimetal phase hosting $N$ gapless Dirac fermions and a long-range-ordered phase which spontaneously breaks the $\mathrm{SO}(3)$ symmetry. Remarkably, this transition differs drastically from the conventional GN case: instead of fully gapping all fermionic excitations, the symmetry-breaking mechanism leaves $N/3$ Dirac cones gapless deep within the ordered phase. Because itinerant fermions remain, this quantum critical point (QCP) represents a Dirac analogue of a metallic QCP, a scenario dubbed metallic GN criticality~\cite{Assaad2022PRL}.

    Despite the exotic critical behavior predicted theoretically, relevant numerical investigations remain scarce. To date, numerical data are only available for $N=12$. In the corresponding honeycomb-lattice model, the $\mathrm{SO}(3)$-broken phase terminates at a putative deconfined quantum critical point (DQCP)~\cite{Assaad2022PRL,Liu2024PRB}. Extended simulations further showed that the fully gapped strong-coupling phase contains symmetry-degenerate interlayer-coherent insulating (ILC) and interlayer $s$-wave superconducting (SSC) orders, related by a partial particle--hole symmetry~\cite{Liu2024PRB}. This raises several compelling questions: How does metallic GN criticality behave for other values of $N$? Is the vanishing of the $\mathrm{SO}(3)$ order universally accompanied by the onset of deconfined quantum criticality? More generally, could other exotic quantum phases emerge in $\mathrm{SO}(3)$-symmetric interacting fermion models? Addressing these questions constitutes a highly intriguing frontier in the field.


    Motivated by these open questions, in this paper, we construct an interacting bilayer model of SLAC fermions with $N=6$ irreducible Dirac cones that respects global $\mathrm{SO}(3)$ symmetry. Upon increasing the interaction strength, we first observe a phase transition breaking $\mathrm{SO}(3)$ symmetry, which belongs to the GN-$\mathrm{SO}(3)$ universality class with $N=6$. We extract the critical point and corresponding critical exponents via quantum Monte Carlo simulations. Upon further enhancing the interactions, the $\mathrm{SO}(3)$ ordered phase terminates abruptly, accompanied by the emergence of an inter-layer superconducting phase. Its pairing structure is equivalent to the SSC component of the $N=12$ strong-coupling manifold~\cite{Liu2024PRB}. The crucial distinction is that the long-range SLAC kinetic term does not preserve the partial particle--hole symmetry that protects the ILC--SSC degeneracy in the honeycomb model; our QMC results instead select superconductivity without accompanying ILC order. Furthermore, this abrupt transition provides strong numerical evidence for a direct first-order boundary rather than a DQCP. Our work thus expands the numerical coverage of metallic GN criticality and demonstrates a distinct, nondegenerate strong-coupling fate of the partially gapped ordered semimetal.



    \begin{figure}[t!]
      \centering
      \includegraphics[width=\linewidth]{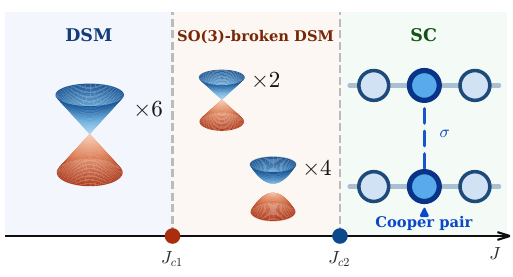}
      \vskip-3mm
      \caption{%
        Schematic ground-state phase diagram as a function of $J$.
        The weak-coupling DSM contains six gapless Dirac cones. In the intermediate $\mathrm{SO}(3)$-broken DSM, four cones are gapped while two remain gapless. At larger coupling, interlayer Cooper pairing produces an SC phase. The transition at $J_{c1}$ is continuous, whereas the abrupt change at $J_{c2}$ strongly indicates a first-order transition.
      }
      \label{figure1}
    \end{figure}

    \begin{figure*}[t!]
      \begin{minipage}[t]{0.475\textwidth}
        \centering
        \includegraphics[width=0.88\linewidth]{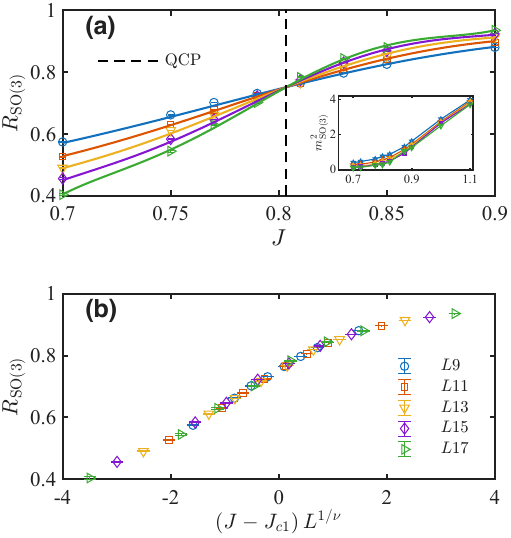}
        \vskip-3mm
        \caption{%
          Correlation ratio $R_{\mathrm{SO}(3)}$ versus $J$ for various sizes (equilibrium). Data points include error bars.
          (a) Estimate of the QCP from the size-independent crossings of $R$. The inset zooms in $m^2_{\mathrm{SO(3)}}$ near $J_{c1}$.
          (b) Scaling collapse of $R$ using the best-fit $J_{c1}$ and $\nu$.
        }
        \label{figure2}
      \end{minipage}
      \hfill
      \begin{minipage}[t]{0.475\textwidth}
        \centering
        \includegraphics[width=0.88\linewidth]{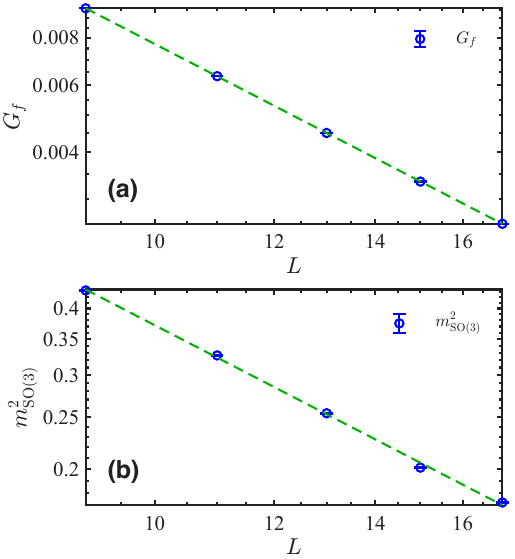}
        \vskip-3mm
        \caption{%
          (a) Log-log plot of the real-space fermion correlation $G_f$ as a function of $L$.
          (b) Log-log plot of the squared order parameter $m^2_{\mathrm{SO}(3)}$ as a function of $L$.%
        }
        \label{figure3}
      \end{minipage}
    \end{figure*}

\PRLsection{Model}
    We consider a bilayer SLAC-fermion model with the Hamiltonian
    \begin{align}
        H={}&\sum_{i,\bm{R},\lambda,\lambda',\sigma}
        \left[c_{i,\uparrow,\lambda,\sigma}^{\dagger}
        (\mathcal T_{\bm{R}})_{\lambda\lambda'}
        c_{i+\bm{R},\downarrow,\lambda',\sigma}+\mathrm{H.c.}\right]
        \nonumber\\
        &-J\sum_{i,\mu,\alpha}
        \left(
        \sum_{\lambda,\lambda',\sigma,\sigma'}
        c_{i,\mu,\lambda,\sigma}^{\dagger}
        K_{\sigma,\sigma'}^{\alpha}
        \tau_{\lambda,\lambda'}^z
        c_{i,\mu,\lambda',\sigma'}
        \right)^2
        \label{eq:Hamiltonian}
    \end{align}
    Here, $c_{i,\mu,\lambda,\sigma}^\dagger$ ($c_{i,\mu,\lambda,\sigma}$) creates (annihilates) a fermion at site $i$, with two-component spin index $\mu=\uparrow,\downarrow$, layer $\lambda=1,2$, and flavor $\sigma=1,2,3$. The two spin components form the Dirac spinor and are coupled by the SLAC hopping term. For a nonzero displacement $\bm{R}=(R_x,R_y)$, the hopping amplitude is
    $t_{\bm{R}}=t\left[\frac{i(-1)^{R_x}}{(L/\pi)\sin(\pi R_x/L)}\delta_{R_y,0}
    +\frac{(-1)^{R_y}}{(L/\pi)\sin(\pi R_y/L)}\delta_{R_x,0}\right]$,
    where $t_{\bm{0}}=0$. Here, $J$ denotes the interaction strength. We set the overall hopping scale $t=1$ and work in two spatial dimensions, $d=2$. The generators act in the three-dimensional flavor space and are normalized as $(K^\alpha)_{\sigma\sigma'}=-i\epsilon_{\alpha\sigma\sigma'}$, whereas the Pauli matrix $\tau^z$ acts in the two-dimensional layer space. SLAC fermion constructions have been used to elucidate Dirac quantum criticality and competing excitonic and superconducting masses in interacting Dirac systems~\cite{Vaezi2022PRL,Li2018SciAdv,Szabo2021JHEP}. The layer-resolved hopping matrix can then be written as $\mathcal T_{\bm{R}}=t_{\bm{R}}P_+ +t_{\bm{R}}^*P_-=\operatorname{Re}(t_{\bm{R}})\tau^0+i\operatorname{Im}(t_{\bm{R}})\tau^z$, where $P_\pm=(\tau^0\pm\tau^z)/2$. Thus the two layers carry conjugate hoppings, $t_{\bm{R}}$ and $t_{\bm{R}}^*$. Consequently, in this model, each layer features a single Dirac fermion per flavor, with opposite chiralities between the two layers.

    Furthermore, the model defined in Eq.~\eqref{eq:Hamiltonian} is free from the fermion sign problem~\cite{Wu2005PRB,Assaad2022PRL,Li2016PRL,Li2015PRB,Xiang2016PRL,Wang2015PRL}, enabling large-scale quantum Monte Carlo simulations. Following a Hubbard--Stratonovich decoupling, the auxiliary-field one-body Hamiltonian remains invariant under the antiunitary transformation $T=i\tau^y\mathcal{K}$, where $\mathcal{K}$ is the complex conjugation operator and $\tau^y$ exchanges the two layers. This symmetry rigorously guarantees the absence of the sign problem, allowing us to perform unbiased simulations on large system sizes. Further technical details are provided in the \hyperref[sec:S_numerical]{Numerical Methods section} of the Supplemental Material. Accordingly, we employ the projector determinant quantum Monte Carlo (DQMC) method to systematically investigate the ground-state properties of this model~\cite{Assaad2002book}.

    Figure~\ref{figure1} summarizes our main results. In the weakly interacting limit, the model features a DSM phase, characterized by $N=6$ irreducible Dirac cones located at the $\Gamma$ point ($\bm{k}=0$). At intermediate couplings, an $\mathrm{SO}(3)$-broken phase emerges, in which two thirds of the Dirac cones are gapped out, while one third remains gapless. Accordingly, this phase is also a DSM phase. For large couplings, the system enters an SC phase.






\PRLsection{$\mathrm{SO}(3)$-broken phase and GN-$\mathrm{SO}(3)$ universality class}
    To reveal the phase transition with $\mathrm{SO}(3)$-symmetry-breaking, we study the structure factor:
    \begin{equation}
      \begin{aligned}
        S(\bm{k})
        &=\frac{1}{L^{2d}}\sum_{i,j,\alpha}
          e^{i\bm{k}\cdot(\bm{r}_i-\bm{r}_j)}
          \left\langle O_i^\alpha O_j^\alpha\right\rangle,\\
        O_i^\alpha
        &\equiv\sum_{\mu,\lambda,\sigma,\sigma'}
          \zeta_\mu\xi_\lambda
          c^\dagger_{i,\mu,\lambda,\sigma}
          K^\alpha_{\sigma\sigma'}
          c_{i,\mu,\lambda,\sigma'}.
      \end{aligned}
        \label{eq:m_SO3}
    \end{equation}
    Here $\zeta_\uparrow=+1$, $\zeta_\downarrow=-1$ and
    $\xi_1=+1$, $\xi_2=-1$, giving the spin- and layer-staggered
    combination used in the simulations.

    The squared order parameter for $\mathrm{SO}(3)$ symmetry breaking is given by $m^2_{\mathrm{SO}(3)} \equiv S(\bm{0})$. As illustrated in the inset of Fig.~\ref{figure2}, $m^2_{\mathrm{SO}(3)}$ exhibits a dramatic increase above $J\approx J_{c1}$, indicating the onset of spontaneous $\mathrm{SO}(3)$ symmetry breaking.


    To pinpoint the QCP and quantitatively resolve the critical features of this phase transition, we employ the dimensionless quantity $R$ constructed from the structure factor $S(\bm{k})$ as follows~\cite{Parisen2015prb}:
    \begin{align}
        R_{\mathrm{SO}(3)} \equiv 1-\frac{S(\Delta \bm{q})}{S(\bm{0})},
        \label{eq:R_SO3}
    \end{align}
    where $\Delta \bm{q}$ is the minimum lattice momentum. The finite-size scaling (FSS) form of $R$ reads
    \begin{align}
        R_{\mathrm{SO}(3)}(g,L)=f_R\!\left(g\,L^{1/\nu}\right),
        \qquad g \equiv J-J_{c1},
        \label{eq:R_eqFSS}
    \end{align}
    with $L$ the linear size, and $\nu$ the correlation-length exponent. According to Eq.~\eqref{eq:R_eqFSS}, at $g=0$ the ratio $R$ becomes size independent; moreover, by tuning $J_{c1}$ and $\nu$, the data for different $L$ collapse onto a single curve near the critical point. These scaling properties of $R$ can be used to determine the critical point and the critical exponent $\nu$.

    Figure~\ref{figure2}(a) shows the dependence of $R_{\mathrm{SO}(3)}$ on $J$ for different sizes $L$. To estimate the critical point and the correlation-length exponent more accurately, we follow the fitting procedure developed in Ref.~\cite{Lang2019PRL}. Near the critical point, the scaling function is approximated by the polynomial
    $f_R\!\left(gL^{1/\nu}\right)=\sum^{n_{\max}}_{n=0}a_n g^nL^{n/\nu}$.
    We fit the data for $L=9,11,13,15,17$ over $0.70\leq J\leq0.90$. Among truncation orders $1\leq n_{\max}\leq5$, the fifth-order polynomial gives the smallest root-mean-square residual. Statistical uncertainties are estimated from the dispersion across 1000 Gaussian-resampled fits. As shown in Fig.~\ref{figure2}(a), the solid curves represent the polynomial fits. Their intersection yields the critical point $J_{c1}=0.803(4)$, indicated by the vertical dashed line, and the correlation-length exponent is $\nu=0.81(5)$. The resulting data collapse of $R_{\mathrm{SO}(3)}$ as a function of $(J-J_{c1})L^{1/\nu}$ is shown in Fig.~\ref{figure2}(b).



     The anomalous dimension for the order parameter, $\eta_b$, can then be determined. At $J=J_{c1}$, $m^2_{\mathrm{SO}(3)}$ satisfies the FSS relation $m^2_{\mathrm{SO}(3)} \propto L^{-1-\eta_b}$. Figure~\ref{figure3}(b) shows the numerical results of $m^2_{\mathrm{SO}(3)}$ at $J_{c1}$ for different $L$. A power-law fit gives $\eta_b=0.44(2)$.

     Similarly, the anomalous dimension of the fermion field, $\eta_f$, can be determined from the channel-resolved correlation $C_{\mu\lambda\sigma}(\bm{R}_m)=L^{-2}\sum_i\langle c^{\dagger}_{i,\mu,\lambda,\sigma}c_{i+\bm{R}_m,\mu,\lambda,\sigma}+\mathrm{H.c.}\rangle$. The quantity used in the finite-size analysis is
    \begin{equation}
        G_f=\frac{1}{12}\sum_{\mu,\lambda,\sigma}
        \left|C_{\mu\lambda\sigma}(\bm{R}_m)\right| .
        \label{eq:G_f}
    \end{equation}
    Here $\bm{R}_m=((L-1)/2,(L-1)/2)$ represents the maximum distance between lattice sites. The absolute value is taken for each of the $2\times2\times3=12$ spin, layer, and flavor channels before averaging, as in the data analysis. At $J_{c1}$, $G_f$ obeys the FSS relation $G_f \propto L^{-2-\eta_f}$. As shown in Fig.~\ref{figure3}(a), a power-law fit to $G_f$ at $J_{c1}$ yields $\eta_f=0.051(2)$.

    Comparing our results with those from renormalization-group (RG) analyses~\cite{Ray2021Fractionalized}, we find $1/\nu=1.23$ to be close to the prediction from the $1/N$ expansion, which yields $1/\nu\approx1.26$. Furthermore, our extracted exponents $\eta_b=0.44(2)$ and $\eta_f=0.051(2)$ are both smaller than the corresponding RG results. Similar discrepancies also appear for $N=12$~\cite{Assaad2022PRL}. This suggests that continued improvements across different theoretical and numerical methods are necessary to achieve consensus.



\PRLsection{SC phase in the strong-coupling region}
For the $N=12$ honeycomb model, a putative DQCP at large coupling separates the $\mathrm{SO}(3)$-broken semimetal from a fully gapped phase whose ILC and SSC components are symmetry-degenerate owing to a partial particle--hole symmetry~\cite{Assaad2022PRL,Liu2024PRB}. We define $D^{\dagger}_{i,\mu,\sigma}=c^{\dagger}_{i,\mu,1,\sigma}c_{i,\mu,2,\sigma}$ for the local inter-layer coherence and use the same spin-staggering factors, $\zeta_\uparrow=+1$ and $\zeta_\downarrow=-1$, as in Eq.~\eqref{eq:m_SO3}. The corresponding squared order parameter is
    \begin{equation}
        m^2_{\mathrm{U}(1)}=\frac{1}{L^{2d}}\sum_{i,j,\mu,\mu',\sigma}
        \zeta_\mu\zeta_{\mu'}\operatorname{Re}\!\left\langle
        D^{\dagger}_{i,\mu,\sigma}D_{j,\mu',\sigma}\right\rangle .
        \label{eq:S_U1}
    \end{equation}

\begin{figure}[t!]
      \centering
      \includegraphics[width=\linewidth]{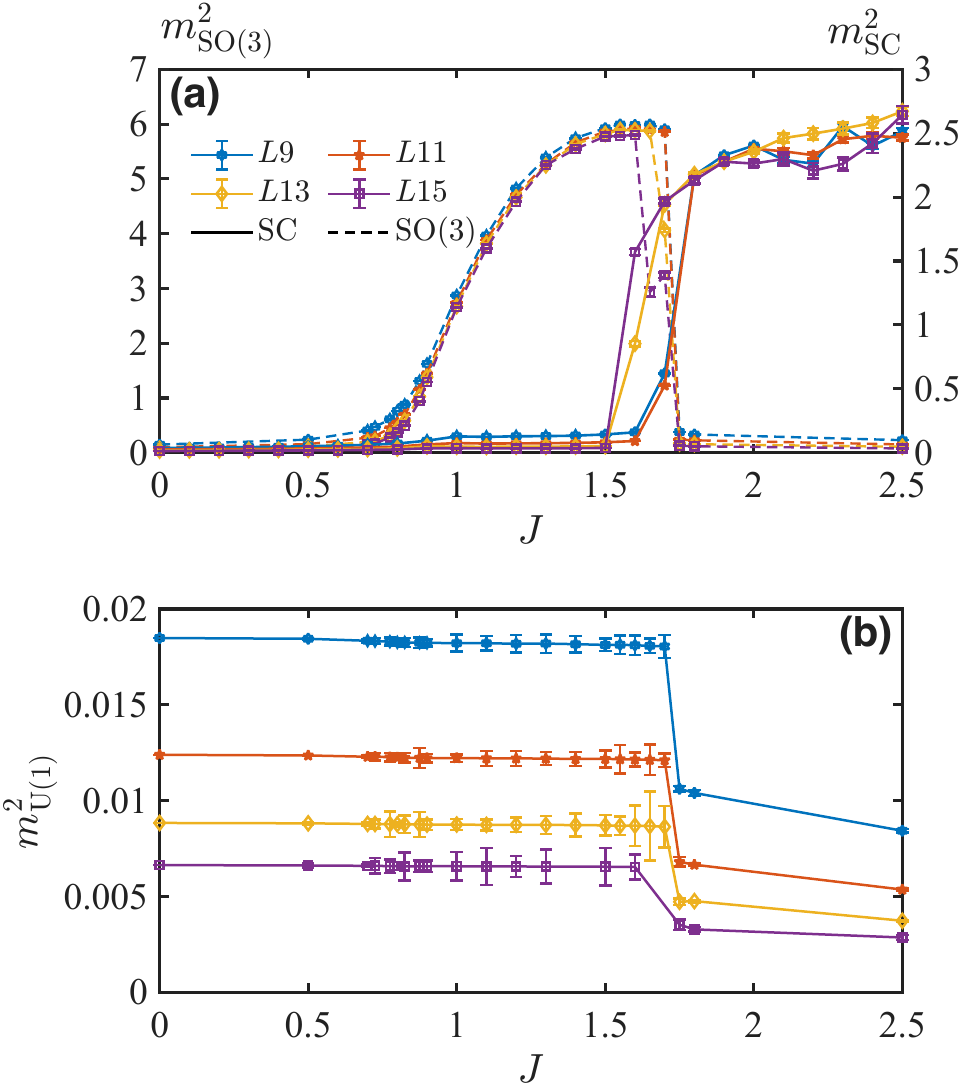}
      \vskip-3mm
      \caption{The order parameters as functions of $J$, where the legend indicates the linear size $L$.
      (a) $m^2_{\mathrm{SO}(3)}$ (left axis, dashed) and $m^2_{\mathrm{SC}}$ (right axis, solid) on a dual-axis plot.
      (b) $m^2_{\mathrm{U}(1)}$ versus $J$ for different sizes.
      }
      \label{figure5}
    \end{figure}

    We therefore examine whether an analogous deconfined transition occurs in the SLAC bilayer. As shown in Fig.~\ref{figure5}(a), $m^2_{\mathrm{SO}(3)}$ drops sharply at large coupling, in stark contrast to the continuous behavior reported for $N=12$~\cite{Assaad2022PRL,Liu2024PRB}. This discontinuity strongly indicates a first-order phase transition out of the $\mathrm{SO}(3)$-broken state. We next measure the ILC order parameter in the SLAC bilayer. Figure~\ref{figure5}(b) shows no ILC long-range order in the strong-coupling phase, ruling out the symmetry-degenerate ILC--SSC manifold realized in the $N=12$ model.

    To identify the phase replacing the $\mathrm{SO}(3)$ state, we first perform an unrestricted Hartree-Fock-Bogoliubov (HFB) calculation, detailed in the \hyperref[sec:S_HFB]{Mean-field analysis} section of the Supplemental Material. This analysis identifies intra-flavor, intra-spin, and inter-layer pairing as the dominant pairing channel. Guided by this, we perform systematic QMC simulations to investigate the nature of the superconductivity. The relevant local pairing operator is defined as
\begin{equation}
    \Delta^{\dagger}_{i,\mu,\sigma} = c^{\dagger}_{i,\mu,1,\sigma}c^{\dagger}_{i,\mu,2,\sigma},
    \label{eq:pair_operator}
\end{equation}
where $\Delta_{i,\mu,\sigma} = (\Delta^\dagger_{i,\mu,\sigma})^\dagger$, with $\mu,\mu'$ denoting the spin indices and $\sigma,\sigma'$ the flavor indices. This operator has the same microscopic inter-layer, intra-flavor, and $\mathrm{SO}(3)$-symmetric structure as the SSC order identified in the $N=12$ model~\cite{Liu2024PRB}. There, partial particle--hole symmetry maps SSC to ILC and enforces their degeneracy. The long-range SLAC kinetic term does not preserve this symmetry, and our QMC results instead select SSC without accompanying ILC order. The corresponding squared order parameter is evaluated as
\begin{equation}
    m^2_{\mathrm{SC}} = \frac{1}{2L^{2d}}\sum_{i,j,\mu,\mu',\sigma,\sigma'} \bigl\langle \Delta^{\dagger}_{i,\mu,\sigma}\Delta_{j,\mu',\sigma'} + \Delta_{j,\mu',\sigma'}\Delta^{\dagger}_{i,\mu,\sigma}\bigr\rangle.
    \label{eq:SC}
\end{equation}

    As depicted in Fig.~\ref{figure5}(a), the sudden onset of $m^2_{\mathrm{SC}}$ at $J_{c2}$ (right vertical axis) is accompanied by a simultaneous drop in $m^2_{\mathrm{SO}(3)}$ (left axis). This coincident discontinuity provides strong numerical evidence for a direct first-order quantum phase transition between the $\mathrm{SO}(3)$-broken phase and the inter-layer $\mathrm{SO}(3)$-symmetric superconducting state.

    To test whether the strong-coupling signal in Fig.~\ref{figure5}(a) is specific to the pairing channel defined in Eq.~\eqref{eq:SC}, we also examine two inter-flavor alternatives. Neither develops an onset comparable to that of $m^2_{\mathrm{SC}}$; their definitions and finite-size behavior are given in the \hyperref[sec:S_competing_SC]{Competing pairing channels section} of the Supplemental Material.

    In summary, our unbiased QMC results demonstrate that an $\mathrm{SO}(3)$-symmetric, intra-flavor, and inter-layer SC order emerges in the strong-coupling regime once the $\mathrm{SO}(3)$-broken phase vanishes. HFB identifies the same leading pairing channel but predicts a continuous onset and a stable region of $\mathrm{SO}(3)$--SC coexistence, whereas QMC reveals a direct first-order transition. This contrast suggests that quantum fluctuations preempt the mean-field coexistence regime~\cite{Calabrese2003,Aharony2003,Janssen2018Compatible,Roy2018Multicritical}. Moreover, the $\mathrm{SO}(3)$ and SC order parameters do not form the conventional quintuplet of mutually anticommuting masses, so the compatible-mass mechanism invoked for the $N=12$ deconfined transition does not directly apply here. Further details are given in the \hyperref[sec:S_HFB]{Mean-field analysis} section of the Supplemental Material.


\PRLsection{Summary and discussion}
    In this work, we studied the ground-state phase diagram and critical phenomena of a bilayer SLAC fermion model with $\mathrm{SO}(3)$-symmetric interactions using projector DQMC simulations. Our study uncovers a continuous phase transition from the DSM to an $\mathrm{SO}(3)$-symmetry-broken DSM phase, which we establish as belonging to the GN-$\mathrm{SO}(3)$ universality class with $N=6$. By systematically characterizing the critical regime, we extracted the corresponding critical exponents. Interestingly, while the correlation length exponent $\nu$ agrees with available analytical calculations, the anomalous dimensions of both the order parameter and the fermion field deviate substantially from RG predictions. Taken together with previous $N=12$ data, these results highlight a notable quantitative discrepancy between numerical and analytical approaches, emphasizing the need for further theoretical and computational efforts to reach a full consensus on this exotic phase transition.

    Additionally, upon increasing the interaction strength, we uncover an abrupt transition from the $\mathrm{SO}(3)$-broken semimetal into an inter-layer, intra-flavor $\mathrm{SO}(3)$-symmetric superconducting state. Our numerical data yield strong evidence that this transition is first-order, while the contrast with HFB suggests that quantum fluctuations preempt a mean-field coexistence regime. Although the pairing structure is equivalent to the SSC component of the $N=12$ strong-coupling manifold~\cite{Liu2024PRB}, the absence of the symmetry-protected ILC--SSC degeneracy in the present SLAC model leads to a distinct, nondegenerate superconducting endpoint and a direct first-order boundary.




    Looking forward, our study offers several promising directions. For example, the critical dynamics of the transition from the Dirac semimetal to the $\mathrm{SO}(3)$-broken semimetal can be investigated. Furthermore, recently developed nonequilibrium scaling approaches may be used to uncover the $\mathrm{SO}(3)$ symmetry-breaking transition in models plagued by the fermion sign problem~\cite{Yu2026PRL,Yu2026SciAdv,Zeng2025PRB,Zeng2025NatCommun}.

    {\bf Acknowledgments}---F. Y. Z. and S. Y. are supported by the National Natural Science Foundation of China (Grant No. 12222515), the Research Center for Magnetoelectric Physics of Guangdong Province (Grant No. 2024B0303390001), the Guangdong Provincial Key Laboratory of Magnetoelectric Physics and Devices (Grant No. 2022B1212010008), the Science and Technology Projects in Guangzhou City (Grant No. 2025A04J5408), and the National Science and Technology Major Project in Quantum Science and Technology (Grant No. 2025ZD0300400). Y.K.Y. and Z.X.L. are supported by the National Natural Science Foundation of China under Grant Nos. 12347107 and 12474146, and Beijing Natural Science Foundation under Grant No. JR25007.  Z.-X.L. is supported by the New Cornerstone Investigator Program.


\bibliographystyle{apsrev4-1}
\bibliography{ref}

\appendix
\onecolumngrid
\newpage
\widetext
\raggedbottom
\thispagestyle{empty}

\setcounter{equation}{0}
\setcounter{figure}{0}
\setcounter{table}{0}
\renewcommand{\theequation}{S\arabic{equation}}
\renewcommand{\thefigure}{S\arabic{figure}}
\renewcommand{\thetable}{S\arabic{table}}
\renewcommand{\theHequation}{S\arabic{equation}}
\renewcommand{\theHfigure}{S\arabic{figure}}
\renewcommand{\theHtable}{S\arabic{table}}

\pdfbookmark[0]{Supplemental Material}{SM}
\begin{center}
    \vspace{3em}
    {\Large\textbf{Supplemental Material for Metallic Gross-Neveu criticality and superconductivity on the $\mathrm{SO}(3)$ SLAC fermion}} \vspace{1em} {\large\textbf{}} \vspace{0.5em}
\end{center}

\section{Numerical Methods}
\label{sec:S_numerical}
    We use projector determinant quantum Monte Carlo to evaluate ground-state
    expectation values~\cite{Assaad2002book} according to
    \begin{equation}
        \langle\hat{A}\rangle
        =
        \frac{
        \langle\Psi_T|e^{-\Theta H/2}\hat{A}e^{-\Theta H/2}|\Psi_T\rangle
        }{
        \langle\Psi_T|e^{-\Theta H}|\Psi_T\rangle
        },
        \label{eq:S_projector}
    \end{equation}
    where $|\Psi_T\rangle$ is a Slater determinant with nonzero overlap with the ground state in the chosen particle-number sector. For all system sizes reported in the main text, we use $\Theta=30$ and $\Delta\tau=0.05$, corresponding to $N_\tau=600$ time slices. On a finite lattice, the three flavor copies of the Dirac cone form a degenerate single-particle shell. An unequal occupation of this shell would explicitly select a flavor direction already in $|\Psi_T\rangle$ and interfere with the subsequent projection. The bilayer trial state is chosen as a pair of conjugate Slater determinants, $|\Psi_T\rangle=|P\rangle_1\otimes|P^*\rangle_2$. To construct $|P\rangle_1$, we diagonalize the noninteracting SLAC hopping matrix in one flavor sector and occupy its lowest $L^2+1$ eigenstates. The same occupied orbitals are assigned to each of the three flavors, giving
    \begin{equation}
        |P\rangle_1
        =
        \prod_{\sigma=1}^{3}\prod_{n=1}^{L^2+1}
        d_{n\sigma}^{\dagger}|0\rangle,
        \qquad
        d_{n\sigma}^{\dagger}
        =
        \sum_{i,\mu}\phi_n(i,\mu)
        c_{i,\mu,1,\sigma}^{\dagger},
        \label{eq:S_trial}
    \end{equation}
    and hence $N_e=3(L^2+1)$ occupied orbitals in each layer determinant. This flavor-balanced closed-shell construction preserves the $\mathrm{SO}(3)$ symmetry of the trial state, avoids an arbitrary resolution of the finite-size Dirac-shell degeneracy, and prevents an artificial flavor bias in the $\mathrm{SO}(3)$ structure factor defined in Eq.~\eqref{eq:m_SO3}.

    We discretize $\Theta=N_\tau\Delta\tau$ and use a symmetric Suzuki--Trotter decomposition. Writing the interaction as
    \begin{equation}
        H_J = -J\sum_{i,\mu,\alpha} O_{i,\mu,\alpha}^2,\qquad
        O_{i,\mu,\alpha} \equiv c^\dagger_{i,\mu}\bigl(K^\alpha\otimes\tau^z\bigr)c^{\phantom\dagger}_{i,\mu},
        \label{eq:S_O}
    \end{equation}
    the short-time propagator is implemented as
    \begin{equation}
        e^{-\Delta\tau H}
        \simeq
        e^{-\Delta\tau H_t/2}
        \left[
        \prod_{i,\mu,\alpha}
        e^{\Delta\tau J O_{i,\mu,\alpha}^2}
        \right]
        e^{-\Delta\tau H_t/2}.
        \label{eq:S_trotter}
    \end{equation}
    The long-range hopping matrix is diagonalized once, so the two kinetic half-step propagators in Eq.~\eqref{eq:S_trotter} are evaluated directly rather than factorized into individual hopping terms. Each interaction factor is decoupled using the two-valued Hubbard--Stratonovich (HS) auxiliary field employed in the simulations,
    \begin{equation}
        e^{\Delta\tau J O^2}
        \simeq
        \frac12\sum_{s=\pm1} e^{\lambda_{\mathrm{HS}} s O},
        \label{eq:S_hs}
    \end{equation}
    with $\lambda_{\mathrm{HS}}=\tfrac12\operatorname{acosh}(e^{4J\Delta\tau})$.
    Equation~\eqref{eq:S_hs} is the finite-time-step discrete HS approximation implemented in the code; it is not treated as an exact identity over the full Fock-space spectrum. We explicitly checked the time-step dependence of the observables and found $\Delta\tau=0.05$ sufficient within their statistical resolution.

    For a given generator $K^\alpha$, only two of the three flavors are mixed. In this active subspace, $K^\alpha$ is unitarily equivalent to $\sigma^y$. The transformation and the resulting HS factor in one layer are
    \begin{equation}
        U_K = \frac1{\sqrt2}\begin{pmatrix} 1 & 1 \\ i & -i \end{pmatrix},
        \quad U_K^\dagger\sigma^y U_K = \sigma^z,
        \qquad
        e^{\lambda_{\mathrm{HS}}s\,\sigma^y}
        =
        U_K
        \begin{pmatrix}
        e^{\lambda_{\mathrm{HS}}s} & 0 \\
        0 & e^{-\lambda_{\mathrm{HS}}s}
        \end{pmatrix}
        U_K^\dagger.
        \label{eq:S_UK}
    \end{equation}
    The sign of $\lambda_{\mathrm{HS}}s$ is reversed in the other layer because of $\tau^z$. At each time slice, these factors are applied to the flavor pairs $(2,3)$, $(3,1)$, and $(1,2)$ for $\alpha=1,2,3$, respectively, and to every site and spin component. Denoting the resulting interaction factors by $V_{\alpha,\tau}$, the one-layer time-slice matrix used in the code is
    \begin{equation}
        B_\tau^{(1)}
        =
        e^{-\Delta\tau h_t/2}
        V_{3,\tau}V_{2,\tau}V_{1,\tau}
        e^{-\Delta\tau h_t/2},
        \label{eq:S_Bslice}
    \end{equation}

    where $h_t$ is the layer-1 hopping matrix and the rightmost flavor factor
    acts first. Eq.~\eqref{eq:Hamiltonian} gives
    $h_t^{(2)}=(h_t^{(1)})^*$; reversing the HS coupling in layer~2 likewise
    complex conjugates each flavor factor because $(K^\alpha)^*=-K^\alpha$.

    Hence, for every auxiliary-field configuration $s$, the two layer
    propagators are complex conjugates and the fermionic weight is
    nonnegative~\cite{Assaad2022PRL}:
    \begin{equation}
        B_\tau^{(2)}[s]
        =
        \bigl(B_\tau^{(1)}[s]\bigr)^*,
        \quad M_2[s]=M_1[s]^*,
        \qquad
        W[s]=\det M_1[s]\det M_2[s]
        =\left|\det M_1[s]\right|^2\geq0.
        \label{eq:S_layer_relation}
    \end{equation}
    Equation~\eqref{eq:S_layer_relation} also means that only one layer has to be propagated explicitly. The second-layer Green functions and observables are reconstructed by complex conjugation. The explicit single-particle basis therefore has dimension $M=2\ \text{(spin)}\times3\ \text{(flavor)}\times L^2=6L^2$, rather than $12L^2$.

    Let $P\in\mathbb{C}^{M\times N_e}$ contain the occupied orbitals of $|\Psi_T\rangle$. The right and left Slater matrices at time slice $\tau$ are
    \begin{equation}
        R_\tau
        =
        B_\tau B_{\tau-1}\cdots B_1P,
        \qquad
        L_\tau
        =
        P^\dagger B_{N_\tau}B_{N_\tau-1}\cdots B_{\tau+1},
        \label{eq:S_prop}
    \end{equation}
    with $R_\tau\in\mathbb{C}^{M\times N_e}$ and $L_\tau\in\mathbb{C}^{N_e\times M}$. The equal-time one-body density matrix and Green function are evaluated from the $N_e\times N_e$ overlap matrix,
    \begin{equation}
        \rho_\tau
        =
        R_\tau(L_\tau R_\tau)^{-1}L_\tau,
        \qquad
        G_\tau=I-\rho_\tau,
        \label{eq:S_G}
    \end{equation}
    The $\mathrm{SO}(3)$, $\mathrm{U}(1)$, and pairing structure factors defined in Eqs.~\eqref{eq:m_SO3}, \eqref{eq:S_U1}, and~\eqref{eq:SC}, respectively, are then evaluated from Wick contractions of these Green functions. A local update $s\to-s$ modifies only the two active flavor channels. If $\rho_{\tau,\mathcal A}$ denotes the corresponding $2\times2$ block of $\rho_\tau$, the update matrix, determinant ratio, and bilayer acceptance probability are
    \begin{equation}
        \Delta_s
        =
        \begin{pmatrix}
        e^{-2\lambda_{\mathrm{HS}}s}-1 & 0\\
        0 & e^{2\lambda_{\mathrm{HS}}s}-1
        \end{pmatrix},
        \qquad
        r_1=\det\!\left(I_2+\Delta_s\rho_{\tau,\mathcal A}\right),
        \quad
        p_{\mathrm{acc}}=\min\!\left(1,|r_1|^2\right).
        \label{eq:S_update}
    \end{equation}
    Thus no full fermion-matrix determinant is recomputed for a local proposal.
    To control the conditioning of the propagated Slater matrices, QR-based
    stabilization~\cite{Sorella1989EPL} is performed every
    $N_{\mathrm{wrap}}=5$ slices. Equal-time observables are measured only at
    the midpoint $\tau=\Theta/2$ of the projection. The initial thermalization
    bins are discarded during post-processing.

\section{Competing pairing channels}
    \label{sec:S_competing_SC}
    The superconducting structure factor defined in Eq.~\eqref{eq:SC} probes intra-flavor inter-layer pairing and develops an abrupt onset at $J_{c2}$ in Fig.~\ref{figure5}(a). To determine whether this response is specific to that channel, rather than a generic enhancement of pairing correlations at large $J$, we compare it with two symmetry-distinct inter-flavor channels. These observables test the channel selectivity of the strong-coupling response and are not used to locate the transition.

    The first channel pairs different flavors across the two layers. We define
    \begin{align}
        m^2_{\mathrm{SC1}} = \frac{1}{2L^{2d}} \sum_{i,j,\mu,\mu',\langle\alpha,\beta\rangle}
        \langle \Delta^{\dagger}_{i,\mu} \Delta_{j,\mu'} +
        \Delta_{i,\mu} \Delta^{\dagger}_{j,\mu'} \rangle,
        \label{eq:SC1}
    \end{align}
    where $\Delta_{j,\mu'}=c_{j,\mu',2,\beta}c_{j,\mu',1,\alpha}$, $\Delta^{\dagger}_{i,\mu}=c^{\dagger}_{i,\mu,1,\alpha}c^{\dagger}_{i,\mu,2,\beta}$, and $\langle\alpha,\beta\rangle=\langle x,y\rangle,\langle y,z\rangle,\langle z,x\rangle$ denotes the three cyclic flavor pairs.

    The second channel pairs different flavors within the same layer. Defining $\Delta^\alpha_{i,\mu,\lambda}=\sum_{\sigma,\sigma'}c_{i,\mu,\lambda,\sigma}K^\alpha_{\sigma\sigma'}c_{i,\mu,\lambda,\sigma'}$, its squared order parameter is
    \begin{equation}
        m^2_{\mathrm{SC2}}
        =\frac{1}{L^{2d}}\sum_{i,j,\mu,\mu',\lambda,\lambda',\alpha}
        (-1)^{\lambda+\lambda'}
        \left\langle\Delta^{\alpha\dagger}_{i,\mu,\lambda}
        \Delta^\alpha_{j,\mu',\lambda'}\right\rangle .
      \label{eq:SC2}
    \end{equation}

    \begin{figure}[t]
      \centering
      \includegraphics[width=0.82\linewidth]{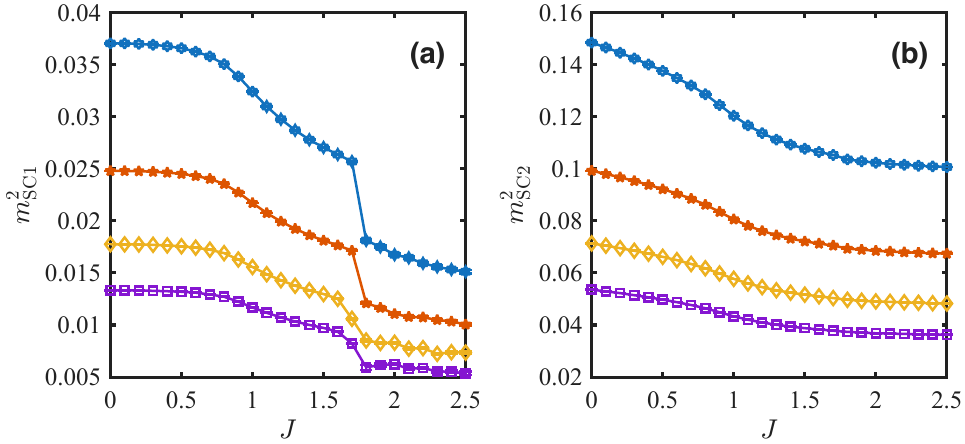}
      \caption{%
        Other pairing channels as functions of $J$.
        (a) $m^2_{\mathrm{SC1}}$ for inter-layer pairing of different flavors.
        (b) $m^2_{\mathrm{SC2}}$ for intra-layer pairing of different flavors.
      }
      \label{fig:S_competing_SC}
    \end{figure}

    Figure~\ref{fig:S_competing_SC} shows that both competing structure factors decrease systematically with increasing $L$ throughout the interaction range. Neither exhibits the abrupt growth at $J_{c2}$ found for $m^2_{\mathrm{SC}}$. Among the three channels compared here, the QMC data therefore single out intra-flavor inter-layer pairing as the strong-coupling instability.

    The following unrestricted HFB analysis examines the pairing channels independently and addresses how the selected superconducting order competes with the $\mathrm{SO}(3)$ mass.
    \FloatBarrier

\section{Mean-field analysis}
    \label{sec:S_HFB}
    The loss of \(\mathrm{SO}(3)\) order at the second transition does not
    identify the large-\(J\) phase.  Since the QMC data exclude the
    symmetry-degenerate ILC--SSC manifold found in the \(N=12\) model
    by showing no accompanying ILC order~\cite{Liu2024PRB},
    we use an unrestricted zero-temperature Hartree--Fock--Bogoliubov (HFB)
    calculation to screen the local particle-hole and pairing channels.  The
    calculation extends the normal-state treatment of
    Ref.~\cite{Assaad2022PRL} to the anomalous sector.

    For the present interaction, the normal and anomalous densities must be
    retained simultaneously in the complete local basis
    \(a=(\mu,\lambda,\sigma)\); no pairing channel is imposed in the
    decoupling.  With
    \(A^{\mu\alpha}=P_\mu\otimes\tau^z\otimes K^\alpha\), where
    \(P_\mu=\lvert\mu\rangle\langle\mu\rvert\) projects onto spin
    component \(\mu\), we retain the full
    complex \(12\times12\) normal and antisymmetric anomalous density matrices,
    denoted by \(\rho\) and \(\kappa\), where
    \(\kappa_{ab}=L^{-2}\sum_{\bm{k}}
    \langle c_{-\bm{k},b}c_{\bm{k},a}\rangle\).
    All self-consistent branches are compared using the expectation value of
    the original Hamiltonian,
    \begin{figure}[t]
        \centering
        \includegraphics[width=0.96\linewidth]{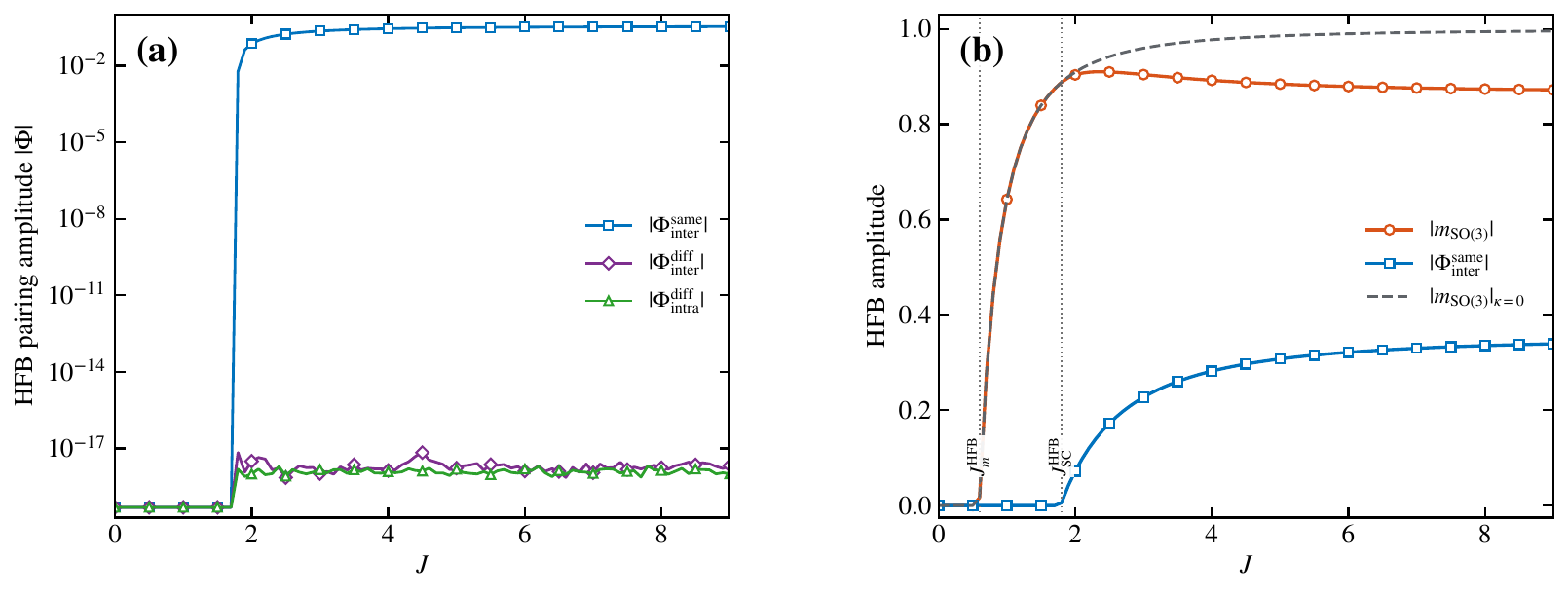}
        \caption{\label{fig:S_HFB}
        Unrestricted HFB results for \(L=71\) with a \((\pi,\pi)\) boundary
        twist.  (a) The three amplitudes defined in
        Eq.~\eqref{eq:S_HFB_pairing_amplitudes}.  Only the same-flavor
        interlayer channel becomes finite; the different-flavor interlayer
        and different-flavor intralayer channels remain below \(10^{-17}\).
        (b) Lowest-energy \(\mathrm{SO}(3)\) and same-flavor interlayer-pairing
        amplitudes.  The dashed curve is the \(\mathrm{SO}(3)\) result when
        pairing is constrained to \(\kappa=0\); the dotted lines mark the two
        HFB onsets.}
    \end{figure}
    \begin{equation}
      \begin{aligned}
        \mathcal E[\rho,\kappa]
        ={}&
        \frac{1}{L^2}\sum_{\bm{k}}
        \operatorname{Tr}\!\left[h_0(\bm{k})\rho(\bm{k})\right]
        -J\sum_{\mu,\alpha}
        \mathcal W_{A^{\mu\alpha}}[\rho,\kappa],
        \\
        \mathcal W_A[\rho,\kappa]
        ={}&
        [\operatorname{Tr}(A\rho)]^2
        +\operatorname{Tr}(A^2\rho)
        -\operatorname{Tr}(A\rho A\rho)
        -\sum_{i,j,k,l}A_{ij}A_{kl}\kappa_{ik}^*\kappa_{lj}.
      \end{aligned}
        \label{eq:S_HFB_energy}
    \end{equation}
    We solve the Bogoliubov--de Gennes equations on finite SLAC momentum grids
    with a \((\pi,\pi)\) boundary twist to remove exact Dirac-shell zero modes.
    We initialize the self-consistency iteration with symmetric,
    \(\mathrm{SO}(3)\), excitonic, pairing, coexistence, and complex random
    states, and continue each solution in both directions in \(J\).  Among
    solutions with residual below \(10^{-10}\), the branch with the lowest
    value of Eq.~\eqref{eq:S_HFB_energy} is retained.

    For comparison with the QMC observables, we project the converged tensors
    onto the relevant order-parameter channels.  We use
    \(s_\uparrow=+1,s_\downarrow=-1\) for the two Dirac-spinor components
    and \(s_1=+1,s_2=-1\) for the two layers.  The \(\mathrm{SO}(3)\) mass
    and equal-\(\mu\) interlayer pairing tensor are
    \begin{align}
        m_\alpha^{\mathrm{HFB}}
        &=
        \frac14\sum_{\mu,\lambda}s_\mu s_\lambda
        \operatorname{Tr}\!\left[
        K^\alpha\rho_{(\mu\lambda),(\mu\lambda)}
        \right],
        \qquad
        \Phi^{\mathrm{inter}}_{\sigma\sigma'}
        =
        \frac12\sum_\mu
        \kappa_{(\mu,1,\sigma),(\mu,2,\sigma')}.
        \label{eq:S_HFB_observables}
    \end{align}
    We denote
    \(\lvert m_{\mathrm{SO}(3)}\rvert
    =(\sum_\alpha\lvert m_\alpha^{\mathrm{HFB}}\rvert^2)^{1/2}\).
    The three pairing amplitudes plotted below are
    \begin{align}
        \lvert\Phi_{\mathrm{inter}}^{\mathrm{same}}\rvert
        &=
        \left[
        \frac13\sum_\sigma
        \lvert\Phi^{\mathrm{inter}}_{\sigma\sigma}\rvert^2
        \right]^{1/2},
        \qquad
        \lvert\Phi_{\mathrm{inter}}^{\mathrm{diff}}\rvert
        =
        \left[
        \frac16\sum_{\sigma\ne\sigma'}
        \lvert\Phi^{\mathrm{inter}}_{\sigma\sigma'}\rvert^2
        \right]^{1/2},
        \nonumber\\
        \lvert\Phi_{\mathrm{intra}}^{\mathrm{diff}}\rvert
        &=
        \left[
        \frac1{12}\sum_{\mu,\lambda}
        \sum_{\sigma<\sigma'}
        \left|
        \kappa_{(\mu,\lambda,\sigma),
        (\mu,\lambda,\sigma')}
        \right|^2
        \right]^{1/2}.
        \label{eq:S_HFB_pairing_amplitudes}
    \end{align}
    The first quantity is the same-flavor, equal-\(\mu\) interlayer channel
    measured in QMC.  The HFB quantities in
    Eq.~\eqref{eq:S_HFB_pairing_amplitudes} are anomalous one-body amplitudes,
    whereas the QMC observables are squared finite-size structure factors;
    their numerical values therefore cannot be compared directly with
    \(m^2_{\mathrm{SC}}\), \(m^2_{\mathrm{SC1}}\), or
    \(m^2_{\mathrm{SC2}}\).

    Only \(\lvert\Phi_{\mathrm{inter}}^{\mathrm{same}}\rvert\) becomes finite
    in Fig.~\ref{fig:S_HFB}(a); the other two amplitudes remain below
    \(10^{-17}\).  The same hierarchy is obtained from
    \(m^2_{\mathrm{SC}}\), \(m^2_{\mathrm{SC1}}\), and
    \(m^2_{\mathrm{SC2}}\) in QMC, identifying the large-\(J\) order as
    same-flavor interlayer pairing.

    At the HFB level, both onsets are continuous and are followed by an
    \(\mathrm{SO}(3)+\mathrm{SC}\) coexistence regime.  Moreover,
    \(\lvert m_{\mathrm{SO}(3)}\rvert\) is reduced relative to the constrained
    \(\kappa=0\) solution after pairing sets in, as shown in
    Fig.~\ref{fig:S_HFB}(b).  The suppression shows that the two orders compete
    already at
    mean-field level.  The coexistence solution remains locally stable up to
    \(J=9\), and the pure \(\mathrm{SO}(3)\) and pure SC branches do not cross
    in energy.

    The mass algebra distinguishes this problem from the usual construction
    of compatible magnetic and superconducting orders.  For a unit vector
    \(\bm n\),
    \(\operatorname{spec}(\bm n\cdot\bm K)=\{-1,0,+1\}\) and
    \((\bm n\cdot\bm K)^2=\mathbb{I}-\lvert\bm n\rangle\langle\bm n\rvert\).
    An \(\mathrm{SO}(3)\) mass therefore gaps only two of the three flavor
    combinations.  The matrices \(K^\alpha\) do not form a mutually
    anticommuting Clifford triplet, so the \(\mathrm{SO}(3)\) and SC orders
    cannot be combined into the conventional compatible
    \(\mathrm{SO}(5)\) mass vector~\cite{Demler2004RMP,Janssen2018Compatible,
    Roy2018Multicritical}.

    In coupled-order-parameter theories, a coexistence regime and a direct
    first-order boundary correspond to different multicritical
    structures~\cite{Calabrese2003,Aharony2003}.  Gapless Dirac fermions can
    modify this structure, and the resulting fixed points depend on whether
    the competing masses are compatible~\cite{Janssen2018Compatible,
    Roy2018Multicritical,HerbutScherer2022,Uetrecht2025}.  In the present
    model, QMC shows an abrupt decrease of the \(\mathrm{SO}(3)\) structure
    factor and a simultaneous increase of the SC structure factor in
    Fig.~\ref{figure5}(a), with no resolved coexistence regime.  This
    discrepancy is consistent with spatial and imaginary-time fluctuations,
    absent from static HFB, preempting the mean-field coexistence regime by a
    direct first-order transition.  The HFB calculation thus identifies the
    leading pairing channel and its competition with \(\mathrm{SO}(3)\)
    order, but does not determine the order of the QMC transition.  The
    direction of this reconstruction is model dependent: in the \(N=12\)
    short-range-hopping model, whose strong-coupling ILC and SSC orders are
    symmetry-degenerate, a strongly first-order mean-field transition instead
    appears direct and continuous in QMC~\cite{Assaad2022PRL,Liu2024PRB}, while
    a spin--charge-flip symmetry produces simultaneous antiferromagnetic and SC criticality
    in another Dirac model~\cite{Liu2022AFSC}.

\end{document}